\documentclass[12pt,preprint]{aastex}
\usepackage{amsmath}
\usepackage{graphics}

\makeatletter

\shorttitle{The Scintillation Velocity of PSR~J1141--6545}
\shortauthors{Ord et al.}
\makeatother

\begin{document}

\title{The Scintillation \\
 Velocity of the Relativistic Binary Pulsar PSR~J1141--6545}

\author{S. M. Ord, M. Bailes and W. van Straten}

\affil{Swinburne University of Technology, Centre for Astrophysics and Supercomputing,
Mail 31, P. O. Box 218, VIC 3122, Australia}

\begin{abstract}

We report a dramatic orbital modulation in the scintillation timescale of the
relativistic binary pulsar J1141--6545 that both confirms
the validity of the scintillation speed methodology and
enables us to derive important physical parameters.
We have determined the space velocity, 
the orbital inclination and even
the longitude of periastron of the binary system, which we find to be in 
good agreement with that obtained from pulse timing measurements.
Our data permit two equally-significant physical interpretations
of the system. The system is either an edge-on binary with
a high space velocity ($\sim 115$~km~s$^{-1}$) 
or is more face-on with a much slower
velocity ($\sim 45$~km~s$^{-1}$). 
We favor the former, as it is more consistent with
pulse timing and the distribution of known neutron star masses.
Under this assumption, the runaway velocity of 115~km~s$^{-1}$ 
is much greater than
is expected if pulsars do not receive a natal kick at birth.
The derived inclination of the binary system is \( 76\pm 2.5^{\circ } \)
degrees, implying a companion mass of 1.01~\( \pm  \)~0.02 M\( _{\odot } \)
and a pulsar mass of 1.29~\( \pm  \)~0.02 M\( _{\odot } \).
Our derived physical parameters
indicate that this pulsar should prove to be an excellent laboratory for
tests of gravitational wave emission.

\end{abstract}


\section{Introduction}

The scintillation observed in the radio wavelength emission from
pulsars is considered to be a result of diffraction of the incident
wave front around small scale irregularities in the interstellar
medium (ISM) (Rickett 1969).  The characteristic timescale for these
scintillation events can be hours, minutes, or even seconds and is
highly dependent upon a number of factors: the distance to the source;
the relative velocities of the source, observer and scattering
material; the observing frequency; and the structure and distribution
of the ISM (Cordes, Pidwerbetsky, \& Lovelace 1986; {Cordes} \& {Rickett} 1998).  It has long been considered possible that
observations of binary pulsars would reveal a modulation in
scintillation velocity due to their orbital motion relative to the
scattering medium (Lyne \& Smith 1982). Such orbital modulation in the
scintillation velocity of a binary pulsar has been significantly
detected only in the circular binary PSR~B0655+64 (Lyne 1984).
Dewey {et~al.} (1988) have reported a weak orbital
modulation of scintillation velocity in observations of PSR~B1855+09 and PSR~B1913+16, but were
unable to constrain any system parameters.

The ideal pulsar for such measurements would scintillate on timescales
much smaller than the orbital period and have enough flux to obtain
high signal-to-noise ratios within the scintillation timescale.  In
addition, the orbital period should be much less than the timescale
for changes in the scintillation parameters due to macroscopic
variations in the structure of the ISM. Finally, the pulsar should
display large variations in its transverse orbital velocity.

The relativistic binary pulsar, PSR~J1141--6545, was recently discovered in the
Parkes multibeam pulsar survey (Kaspi {et~al.} 2000). It has a spin period (\( P \))
of 394~ms, a very narrow (\( 0.01P \)) pulse, a dispersion measure of 116 pc~cm$^{-3}$ and is in an eccentric,
4.7~hr orbit. It is thought to be a member of a new class of object
that has a young neutron star in an eccentric orbit with a massive $\sim$1 M$_{\odot}$\,
white dwarf companion. Kaspi {et~al.} (2000) found no evidence for
scintillation. We have searched specifically for short--timescale
scintillation and found that the pulsar scintillates on timescales of
minutes with a characteristic bandwidth of just $\sim$1 MHz at a central observing frequency of 1390~MHz.  At the
Parkes 64\,m observatory it completes two orbits during its transit time,
changes velocity by over 200~km~s$^{-1}$, and has a short orbital
period.  In every respect, this pulsar is an ideal target for
scintillation work.  Recent neutral Hydrogen observations 
(Ord, Bailes \& van Straten 2002)\nocite{obs02} place the binary
at least as distant as the Carina--Sagittarius spiral arm, which is 3.7~kpc
away in the direction of PSR~J1141--6545.

Unlike their progenitors, the massive O and B stars, radio pulsars
have very large space velocities of up to 1000~km~s$^{-1}$ 
(Lyne, Anderson, \& Salter 1982; Bailes {et~al.} 1989; Harrison, Lyne, \& Anderson 1992).  The origin of pulsar velocities is difficult to
ascertain as individual pulsars give little clues regarding the
responsible physical mechanism (Radhakrishnan \& Shukre 1986; Dewey \& Cordes 1987; Bailes 1989).  Eccentric
binary pulsars, on the other hand, are fossil records of the state of
a binary before detonation of the pulsar progenitor.  In the absence
of a natal kick, a binary pulsar has a well-defined relationship
between its eccentricity, the amount of mass ejected during the
formation of the neutron star, and the runaway
velocity (Radhakrishnan \& Shukre 1986). However, eccentric binary pulsars are rare and it
has not been possible to make a definitive statement about kicks from
an analysis of their space velocities and orbital
configurations (Cordes \& Wasserman 1984; Hughes \& Bailes 1999; Wex, Kalogera, \& Kramer 2000). 
Kaspi {et~al.} (1996) have shown that the massive binary PSR
J0045--7319 has a precessing orbit consistent with the pulsar
receiving an impulsive kick at birth. In the case of PSR J1141--6545,
the space velocity, had the pulsar received no natal
kick, should be small as the eccentricity is low. 
But evolutionary arguments concerning the minimum diameter
of the progenitor ({Tauris} \& {Sennels} 2000) suggest that its
runaway velocity should be at least 150~km~s$^{-1}$.

In this paper, we present the detection of an orbital modulation of
the scintillation velocity in observations of PSR~J1141--6545. The
effect is very significant, permitting the determination of a number
of system parameters.  Measurements have been obtained of the most
probable space velocity and inclination of this system, as well
as the longitude of periastron.  As with the analysis of PSR~B0655+64 by
Lyne (1984), there is a clear degeneracy between two distinct
solutions of equal significance.  In this case, however, the
degeneracy can be broken: one of the solutions indicates an
unreasonably low neutron star mass.
The structure of this paper is as follows:  in section 2 we describe
our observations and methodology for determining the scintillation
parameters;  section 3 describes the model and our fitted
results; finally, in section 4 we discuss the results and their
implications for future relativistic observables and the origin of
pulsar velocities.

\section{The Observations and Data Reduction}

PSR J1141--6545 was observed on the 27th of January 2002 for
approximately ten hours, using the Parkes 64\,m radio telescope.  The
\( 512\times0 .5 \) MHz filterbank centred at 1390 MHz was used with
the centre element of the Parkes multi-beam receiver 
(Staveley-Smith {et~al.} 1996).  In each filterbank channel, the
power in both linear polarisations was detected and summed before
1~bit sampling the total intensity every 250 $\mu$s.  The results were
written to magnetic tape for offline processing.

\begin{figure}
{\par\centering \resizebox*{1\textwidth}{0.5\textheight}{\includegraphics{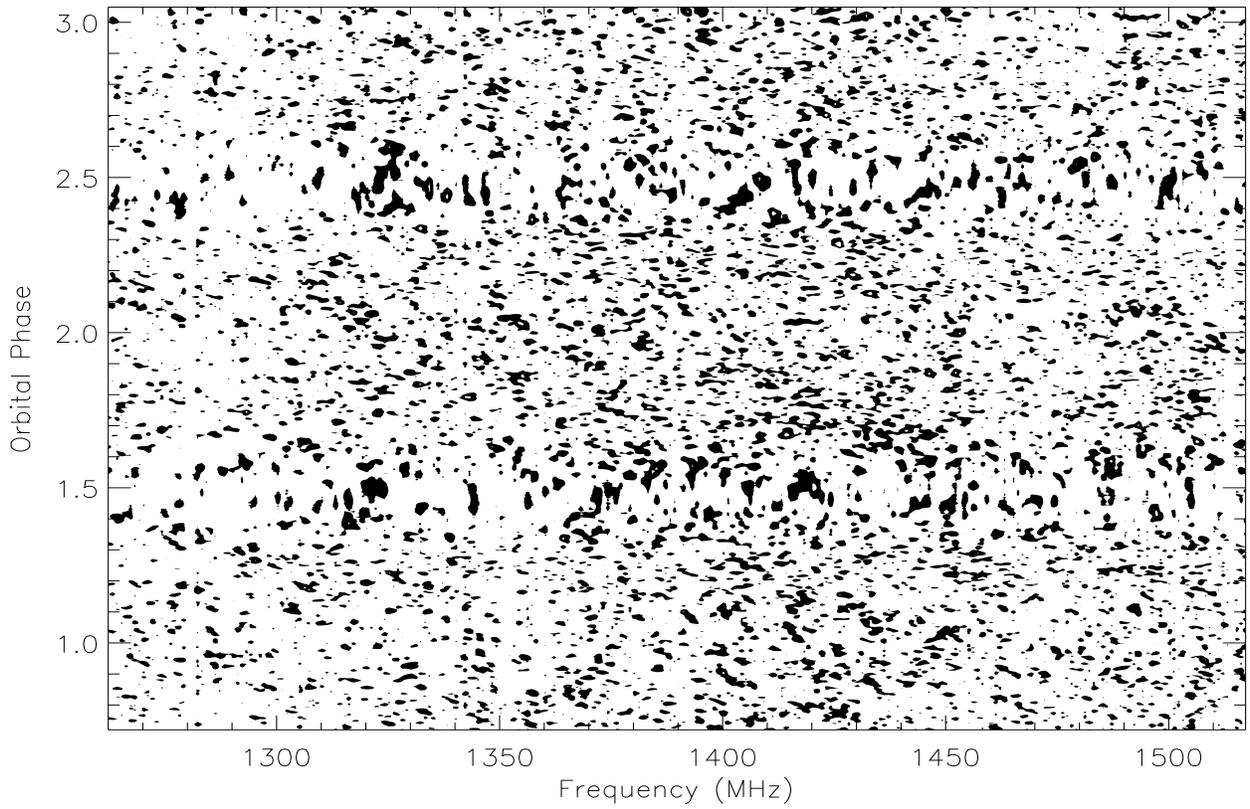}} \par}

\caption{ The dynamic spectrum for PSR J1141--6545 represents the pulsar flux
as a function of time and radio frequency. The plot is presented as a
two level grey scale for emphasis. The vertical blurring
once per orbit is interpreted as a lower relative speed in the
plane of the sky. }
\label{dynamic}
\end{figure}

Average pulse profiles were produced by folding the raw data in each
500 kHz frequency channel modulo the topocentric pulse period.  An
integration time of 29 seconds provided sufficient resolution to
resolve the scintillation structure throughout the orbit.  For each
profile, the mean level of the off-pulse region was subtracted from
that of the on-pulse region to produce a dynamic spectrum.  Figure
\ref{dynamic} shows the clear orbital modulation of the dynamic
spectrum.  In each orbital period, there is an obvious apparent
vertical ``blurring'' as the scintillation timescale varies.

In order to model the physical parameters of the binary system, we
require estimates of the pulsar's velocity as a function of orbital
phase.  Individual measurements of the transverse velocity were
obtained at different epochs throughout the observation by determining
the scintillation timescale and bandwidth.  These were derived from
the two-dimensional auto-correlation function (ACF).

Individual ACFs were calculated over a small region of the dataset
that spanned approximately 12 minutes in the time domain and 24 MHz in
the frequency domain.  A two-dimensional fast Fourier transform (FFT)
was calculated and multiplied by its complex conjugate to form the
ACF.  To improve the signal-to-noise ratio, a single ACF was formed at
each timestep by combining all the ACFs produced across the bandpass.
The scintillation timescale was defined to be the $1/e$ full width of a Gaussian
fit to the central peak of the ACF in the direction of increasing
time-lag.  Following convention (Cordes~1986), the scintillation bandwidth was defined to be half the 
width at half-height of a Gaussian fit across the frequency lags.

By moving the ACF window through the dynamic spectrum with a timestep
of one time lag, or 29 seconds, scintillation parameters were assigned a
time corresponding to the central lag of the ACF.  These values were
integrated into 32 binary phase bins using the ephemeris presented by
Kaspi {et~al.} (2000).  The value of the scintillation velocity was
determined using the following relationship.

\begin{equation}
\label{viss}
V_{\rm ISS}=2.53\times 10^{4}\frac{\left( D\Delta \nu _{\rm d}\right) ^{1/2}}{f\tau _{\rm d}}
\end{equation}

Equation \ref{viss} is after Cordes and Rickett (1998)\nocite{cr98}. \(V_{\rm
ISS} \) is the scintillation velocity, \( D \) is the
earth-pulsar distance in kpc, \( \Delta\nu _{\rm d} \) is the
scintillation bandwidth in MHz and \( \tau _{\rm d} \) is the
scintillation timescale in seconds. The constant multiplicative
factor is the value determined by Cordes and Rickett (1998) for
a uniform Kolmogorov scattering medium. The distance used was
3.7~kpc (Ord, Bailes \& van Straten 2002)\nocite{obs02}.  Any
uncertainty is incorporated into a scaling parameter within the
model.


This process produced the scintillation velocity as a function of mean
orbital anomaly.  The system is significantly eccentric ($e\sim0.17$)
and the model described in \S \ref{model} constructs the velocity as a
function of true, not mean, orbital anomaly.  We therefore transform
the observed scintillation velocity into a function of eccentric
anomaly, $\eta$, by an iterative solution of Kepler's equation,
$\epsilon =\eta +e\sin {\eta }$, where \( \epsilon \) is the mean
anomaly.  Eccentric anomaly is converted into true anomaly, $\theta$,
via a simple trigonometric relationship:
\begin{equation}
\label{trig}
\tan {\frac{{\theta }}{2}}=\sqrt{{{\frac{{1+e}}{1-e}}}}\tan {\frac{{\eta }}{2}}.
\end{equation}

\section{The Model}

\label{model} To determine the most probable runaway velocity and inclination
angle of this system, we constructed a model that calculated the
transverse velocity as a function of five free parameters.  These were:
the components of the transverse velocity 
along and perpendicular to the
\textit{line of nodes}, \( v_{\rm plane} \), and 
\( v_{\rm per} \), respectively; the orbital inclination angle,
$i$; a scaling factor, $\kappa$; and the longitude of periastron, $\omega$.
The latter is accurately determined by pulse timing, and allows us
to confirm the validity of our results.

The orbital velocity was calculated as a function of true orbital
anomaly \( (\theta ) \) and broken into a radial component directed
toward the focus of the orbital ellipse in the line of nodes (\( v_{\rm r}
\)), the second component was perpendicular to \( v_{\rm r} \) and in the
direction of the orbital motion, \( v_{\theta } \).

In this framework, the space velocity as a function of true orbital anomaly
is given by: 
\begin{eqnarray}
\textstyle v_{\rm r} & = & \frac{{2\pi xc}}{\sin i(1-e^{2})^{1/2}P_{\rm b}}e\sin {\theta }\\
v_{\theta } & = & \frac{2{\pi }xc}{\sin i(1-e^{2})^{1/2}P_{\rm b}}(1+e\cos {\theta })
\end{eqnarray}
where the observable $x=(a/c) \sin i$ is the 
projected semi-major axis in seconds,
\( a \) is the semi-major axis of orbit, \( i \)
is the inclination of the system, \( e \) is the orbital eccentricity and 
\( P_{\rm b} \)
the binary period. 

After the construction of the binary model, the next stage
was to transform this orbital velocity into a predicted scintillation velocity.

Assuming that the Earth had a near constant velocity throughout the observation
and that the interstellar medium velocity was small compared to that of the
pulsar, the model of the transverse velocity ($V_{\rm model}$) was described by
the following equations: 
\begin{eqnarray}
v^{\prime } & = & \left( \left( v_{r}\cos {\phi }-v_{\theta }\sin {\phi }\right) +v_{\rm plane}\right) ,\\
v^{\prime \prime } & = & \left( \left( v_{\theta }\cos {\phi }+v_{r}\sin {\phi }\right) \cos {i}+v_{\rm per}\right) ,\\
V_{\rm model} & = & \kappa \sqrt{v^{\prime 2}+v^{\prime \prime 2}}.
\end{eqnarray}
 The angle \( \phi=\omega+\theta  \) is the true anomaly measured with respect
 to the line of nodes.  The model parameter \( \kappa  \) was included to
 incorporate any errors introduced in the conversion of scintillation timescale
 to velocity into the model.  This accounts for errors in the distance to the
 pulsar and the fact that the scattering medium is not uniform. The model was
 then evaluated at the same 32 values of true anomaly presented by the dataset.
 The best fit to the model was found by evaluating the chi-squared statistic
 throughout a wide range of each of the model
 parameters. This produced a chi-squared volume of 5 dimensions with a global
 minimum representing the most probable values of the 5 parameters. 

A box-car smoothing operation was also
applied to the model in order to compensate for the averaging introduced by
the auto-correlation process. This 
had little effect on the values of the fitted parameters
but did appreciably lower the chi-squared minimum. 

The relative errors in each binary phase bin were calculated by first
determining the scatter in each phase bin.  The best fit to
the data was found by location of the global minimum in the
chi-squared space, and the error bars were normalised to ensure the reduced
chi-squared statistic was unity. This enabled errors to be placed upon the
estimates of the fitted parameters by examining the chi-squared
distribution. It should be noted that this method assumes
\textit{a priori} that the model was a valid description of the data. 
A Monte Carlo analysis was then performed in order to ascertain the precision
of the fitted parameters. This analysis demonstrated that
the model described the observations particularly well and the range
of allowable parameter values was very narrow.

\begin{figure}
{\par\centering \resizebox*{1\textwidth}{0.5\textheight}{\rotatebox{270}{\includegraphics{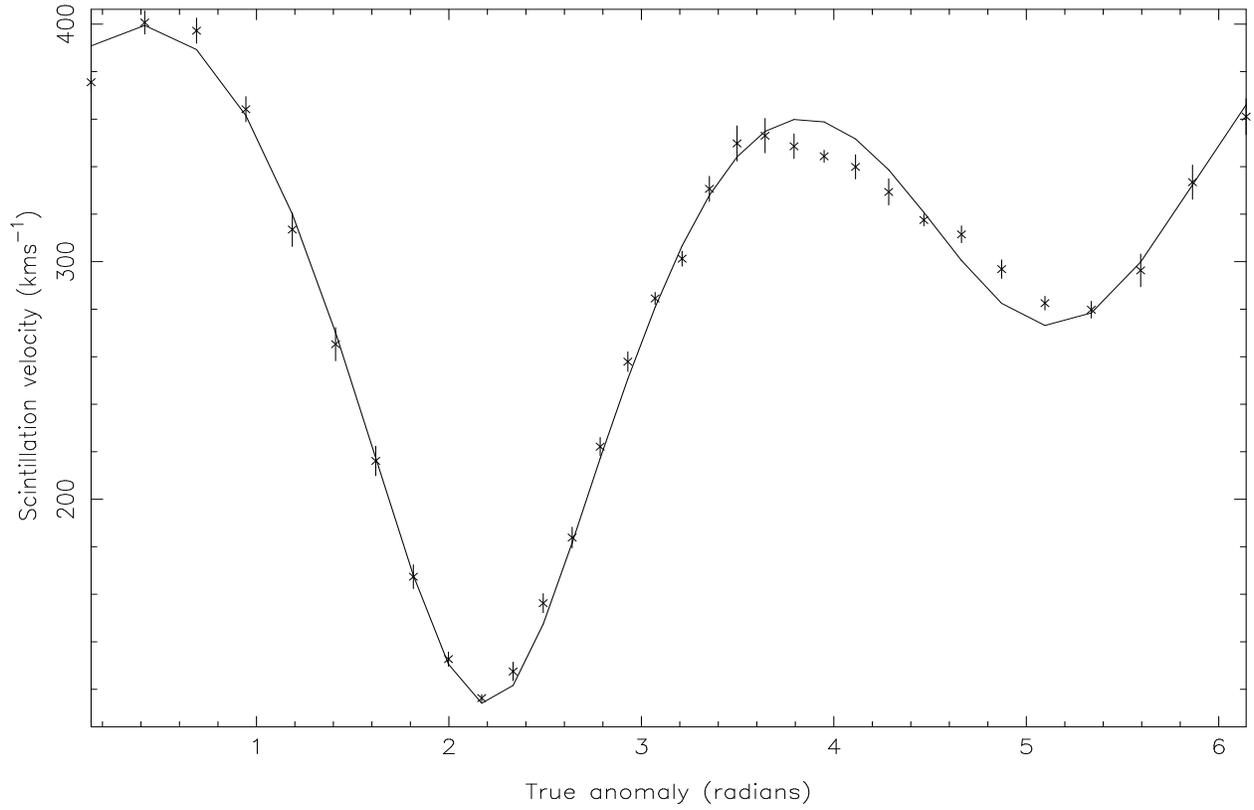}}} \par}

\caption{\label{figviss} -- A plot of scintillation velocity versus
true anomaly for the relativistic binary pulsar, PSR~J1141--6545.  The
solid line represents the best-fit model.  Velocity estimates are plotted with
their one sigma errors, as determined using a $\chi^2$ normalization technique.}
\end{figure}

Both the measured values of \( V_{\rm ISS} \) and the best-fit model velocity, \( V_{\rm model} \), are
plotted as a function of true anomaly in Fig \ref{figviss}.

\subsection{The effects of Refractive Interstellar Scintillation}

The analysis described above would not have revealed any underlying systematic
error introduced by either random or long-term variations in
the scintillation timescale that are not associated with changes  
in the velocity of the pulsar.
For example, the
interstellar scintillation effects detailed here are a result of
diffractive interstellar scintillation (DISS), but another regime of
scintillation is refractive interstellar scintillation (RISS). RISS,
 which has a characteristic timescale generally much longer than
that of DISS, was considered by Romani, Narayan and Blandford
(1986)\nocite{rnb86} to be the explanation for the long term flux
variability in pulsars observed by Sieber
(1982)\nocite{sie82}. Although the effect of RISS is best
characterised via multiple flux measurements, it is possible to
estimate the RISS timescale, \( T_{\rm RISS} \), using the measured DISS
parameters:
\begin{eqnarray}
T_{\rm RISS} & =150.6 & \left( \frac{D({\rm kpc})}{\Delta \nu 
({\rm kHz})V_{7}^{2}}\right) ^{1/2}{\rm days.}\label{Riss} 
\end{eqnarray}

Equation \ref{Riss} is after Blandford and Narayan
(1985)\nocite{bn85}. It assumes a uniform Kolmogorov medium and that
the pulsar is travelling at \( 100 V_{7} \)\,km~s\( ^{-1} \). In Equation \ref{Riss} \( D \)
represents distance and \( \Delta \nu \) is the scintillation
bandwidth. If the velocity of the source can be approximated by the
measured scintillation velocity, then \( T_{\rm RISS} \) is approximately
7 days. It is therefore unlikely that a modulation of this nature was
present in the observation. Even if it was present, the modulation 
would be weak as RISS effects are considerably weaker
than DISS events (Rickett 1990\nocite{ric90}). For these reasons, the
effect of RISS was discounted in this analysis. Future observations,
if they span multiple days, 
may be prone to RISS effects that limit the precision of
derived parameters.

\section{Results}

\label{omega} 
\begin{figure}
{\par\centering \resizebox*{1\textwidth}{0.5\textheight}{\includegraphics[angle=270]{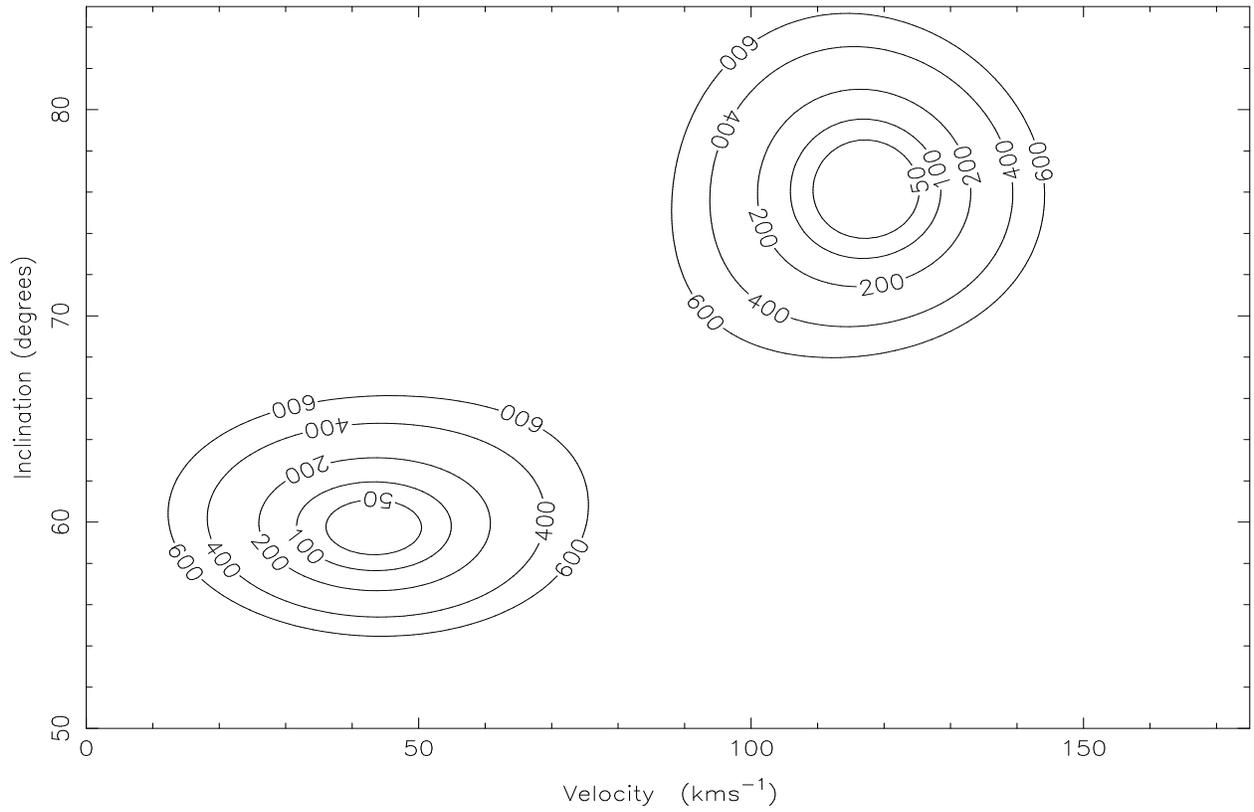}} \par}

\caption{The projection of the delta-chi-squared volume into the dimensions of orbital inclination and transverse velocity. 
The two distinct solutions are evident. }
\label{surface}
\end{figure}

As discussed by Lyne (1984)\nocite{lyn84},   analyses of this
nature can produce two indistinguishable solutions. The degeneracy arises as
the same apparent transverse velocity can be displayed by the pulsar in either
a comparatively face-on binary system with a low runaway velocity, or a more
edge-on system with a higher runaway velocity. An example of this can be seen
in the chi-squared projection presented in Figure \ref{surface}.
The values of the fit parameters for both solutions after the Monte Carlo error
analysis are presented in Table \ref{params}.  The degeneracy between solutions
can be broken by a reasonable consideration of the mass function and other
timing parameters.  Kaspi {et~al.} (2000) derive a value for the mass function,
\( f(M) \) = 0.176 M\(_{\odot } \), and an estimation of the sum of the system
component masses, \( M_{c}= \) 2.3 M\( _{\odot } \), from timing
measurements. Using these measurements, together with the system
inclination presented here, the component masses may be obtained. The
more edge-on solution indicates a pulsar and companion mass of 1.29 \(
\pm  \) 0.02 M\( _{\odot } \) and 1.01 \( \pm  \) 0.02
M\( _{\odot } \) respectively; whereas the more face-on
solution, suggests a pulsar mass of 1.17 \( \pm  \) 0.02 M\( _{\odot } \) and a
companion mass of 1.13 \( \pm  \) 0.02 M\( _{\odot } \). The edge-on solution
is more likely as the indicated pulsar mass is more consistent with the known
neutron star mass distribution, $1.35 \pm 0.04$~M$_{\odot}$ (Thorsett \& Chakrabarty 1999). The most
probable transverse velocity is therefore found to be of modulus 115~\( \pm
\)~15~km s\( ^{-1} \), and the most likely inclination angle is
76~\( \pm  \)~2.5\( ^{\circ } \) .

In order to test the validity of the experimental method employed and the accuracy
of the binary model, we allowed the angle of periastron to vary as a free parameter
in the fit. The value of \( \omega  \) is already precisely determined by timing
measurements and its value at the epoch of these observations is expected to
be approximately 55.13\( ^{\circ } \). The consistency of this with our measurement
of 58 \( \pm  \)3.5\( ^{\circ } \) is very encouraging and
increases our confidence in the solution which 
is extremely statistically significant.



\section{Discussion}

PSR J1141--6545 has been shown to have a transverse velocity of
modulus 115~\( \pm \)~15~km\,s\( ^{-1} \) and an inclination of 76 \( \pm
\) 2.5\( ^{\circ } \). The component masses inferred from these values
indicate, as first presented by Kaspi {et~al.} (2000), 
that this system is most likely a near edge-on
neutron star--CO white dwarf binary, with the observed neutron star
having been formed most recently.

Tauris and Sennels (2000)\nocite{ts00} have predicted, from
evolutionary arguments, that PSR~J1141--6545 would display a systemic
velocity in excess of 150~km\,s\( ^{-1} \).  This velocity comes from
both the recoil of the binary due to the rapid ejection of the
exploding star's envelope, and an impulsive kick imparted to the
neutron star to leave it in an orbital configuration resembling the
current state of PSR~J1141--6545.  The ``no-kick'' solution, suggests
a velocity nearer 40~km~s$^{-1}$.  The measured transverse velocity is
only consistent with that predicted by Tauris \& Sennels (2000), if
the undetected radial velocity exceeds 96~km\,s\( ^{-1} \).
The simulations undertaken by Tauris \& Sennels (2000) suggested
that 150~km~s$^{-1}$ was the lower limit, and that PSR J1141--6545
would probably have a velocity much greater than this. 
The low tranverse velocity indicates that, as with the other
eccentric binary pulsars, the kick was modest (Hughes \& Bailes 1999).

It is also interesting to note that Ord, Bailes \& van Straten (2002)
give a lower distance limit to this system as the tangent point
distance, 3.7~kpc.  At this distance, and with a transverse velocity
of 115~\( \pm \)~15~km~s\( ^{-1} \), the expected contribution to any
measured orbital period derivative due to proper motion and Galactic
kinematic effects is expected to be at the 1 percent level.
Furthermore, measurements of the range and shape of Shapiro delay,
advance of periastron, gravitational redshift parameter and
gravitational period derivative will over-determine the system
parameters. Together with the determination of the system inclination
presented here, estimates of these parameters will allow precise and
unique tests of the validity of General Relativity.

This result suggests that long-term 
monitoring of its scintillation properties throughout a year
may ultimately provide the orientation of the
proper motion vector on the sky. An independent
measure of the orbital period, eccentricity, longitude of
periastron and
inclination angle of the binary should also be obtained.

\bibliography{}

\clearpage

\begin{thebibliography}{}
 
\bibitem[Bailes 1989]{bai89}
Bailes, M. 1989, Astrophys. J., 342, 917
 
\bibitem[Bailes, Manchester, Kesteven, Norris, \&  Reynolds 1989]{bmk+89}
Bailes, M., Manchester, R.~N., Kesteven, M.~J., Norris, R.~P., \& Reynolds,  J.~E. 1989, Astrophys. J., 343, L53
 
\bibitem[Blandford \& Narayan 1985]{bn85}
Blandford, R.~D. \& Narayan, R. 1985, Mon. Not. R. astr. Soc., 213, 591
 
\bibitem[Cordes, Pidwerbetsky, \& Lovelace 1986]{cpl86}
Cordes, J.~M., Pidwerbetsky, A., \& Lovelace, R. V.~E. 1986, Astrophys. J.,  310, 737
 
\bibitem[{Cordes} \& {Rickett} 1998]{cr98}
{Cordes}, J.~M. \& {Rickett}, B.~J. 1998, \apj, 507, 846
 
\bibitem[Cordes \& Wasserman 1984]{cw84}
Cordes, J.~M. \& Wasserman, I. 1984, Astrophys. J., 279, 798
 
\bibitem[Dewey \& Cordes 1987]{dc87}
Dewey, R.~J. \& Cordes, J.~M. 1987, Astrophys. J., 321, 780
 
\bibitem[Dewey, Cordes, Wolszczan, \& Weisberg 1988]{dcww88}
Dewey, R.~J., Cordes, J.~M., Wolszczan, A., \& Weisberg, J.~M. 1988, in Radio  Wave Scattering in the Interstellar Medium ,{AIP} Conference Proceedings  {N}o. 174, ed. J.~Cordes, B.~J. Rickett, \& D.~C. Backer (New York: American  Institute of Physics), 217--221
 
\bibitem[Harrison, Lyne, \& Anderson 1992]{hla92}
Harrison, P.~A., Lyne, A.~G., \& Anderson, B. 1992, in {X}-ray Binaries and  Recycled Pulsars, ed. E.~P.~J. van~den Heuvel \& S.~A. Rappaport (Dordrecht:  Kluwer), 155--160
 
\bibitem[Hughes \& Bailes 1999]{hb99}
Hughes, A. \& Bailes, M. 1999, Astrophys. J., 522, 504
 
\bibitem[Kaspi, Bailes, Manchester, Stappers, \&  Bell 1996]{kbm+96}
Kaspi, V.~M., Bailes, M., Manchester, R.~N., Stappers, B.~W., \& Bell, J.~F.  1996, Nature, 381, 584
 
\bibitem[Kaspi, Lyne, Manchester, Crawford, Camilo, Bell,  D'Amico, Stairs, McKay, Morris, \& Possenti 2000]{klm+00a}
Kaspi, V.~M., Lyne, A.~G., Manchester, R.~N., Crawford, F., Camilo, F., Bell,  J.~F., D'Amico, N., Stairs, I.~H., {et al.}, 2000, Astrophys. J., 543, 321
 
\bibitem[Lyne 1984]{lyn84}
Lyne, A.~G. 1984, Nature, 310, 300
 
\bibitem[Lyne, Anderson, \& Salter 1982]{las82}
Lyne, A.~G., Anderson, B., \& Salter, M.~J. 1982, Mon. Not. R. astr. Soc., 201,  503
 
\bibitem[Lyne \& Smith 1982]{ls82}
Lyne, A.~G. \& Smith, F.~G. 1982, Nature, 298, 825
 
\bibitem[{Ord}, {Bailes}, \& {van Straten} 2002]{obs02}
{Ord}, S.~M., {Bailes}, M., \& {van Straten}, W. 2002, Mon. Not. R. astr. Soc.,  submitted
 
\bibitem[Radhakrishnan \& Shukre 1986]{rs86}
Radhakrishnan, V. \& Shukre, C.~S. 1986, Astrophys. Space Sci., 118, 329
 
\bibitem[Rickett 1969]{ric69}
Rickett, B.~J. 1969, Nature, 221, 158
 
\bibitem[Rickett 1990]{ric90}
Rickett, B.~J. 1990, Ann. Rev. Astr. Ap., 28, 561
 
\bibitem[Romani, Narayan, \& Blandford 1986]{rnb86}         
Romani, R.~W., Narayan, R., \& Blandford, R. 1986, Mon. Not. R. astr. Soc.,  220, 19
 
\bibitem[Sieber 1982]{sie82}
Sieber, W. 1982, Astr. Astrophys., 113, 311
 
\bibitem[Staveley-Smith, Wilson, Bird, Disney,  Ekers, Freeman, Haynes, Sinclair, Vaile, Webster, \& Wright 1996]{swb+96}
Staveley-Smith, L., Wilson, W.~E., Bird, T.~S., Disney, M.~J., Ekers, R.~D.,  Freeman, K.~C., Haynes, R.~F., Sinclair, M.~W., {et al.}, 1996, Proc. Astr. Soc. Aust., 13, 243
 
\bibitem[{Tauris} \& {Sennels} 2000]{ts00}
{Tauris}, T.~M. \& {Sennels}, T. 2000, Astr. Astrophys., 355, 236
 
\bibitem[Thorsett \& Chakrabarty 1999]{tc99}
Thorsett, S.~E. \& Chakrabarty, D. 1999, Astrophys. J., 512, 288
 
\bibitem[Wex, Kalogera, \& Kramer 2000]{wkk00}
Wex, N., Kalogera, V., \& Kramer, M. 2000, Astrophys. J., 528, 401
 
\end{thebibliography}

\clearpage

\begin{table}
{\centering \begin{tabular}{ccc}
\hline 
System Parameter&
Solution 1&
Solution 2\\
\hline 
\hline 
\( \kappa  \) &
1.55 \( \pm  \) 0.025&
1.55 \( \pm  \) 0.025\\
\hline 
 \( v_{\rm plane} \) &
15 \( \pm  \) 10 km s\( ^{-1} \) &
20 \( \pm  \) 10 km s\( ^{-1} \) \\
\hline 
 \( v_{\rm per} \) &
115 \( \pm  \) 10 km s\( ^{-1} \)&
40 \( \pm  \) 10 km s\( ^{-1} \)\\
\hline 
 \( i \) &
76 \( \pm  \) 2.5\( ^{\circ } \) &
60 \( \pm  \) 2.5\( ^{\circ } \) \\
\hline 
 \( \omega  \) &
58 \( \pm  \) 3.5\( ^{\circ } \) &
58 \( \pm  \) 3.5\( ^{\circ } \) \\
\hline 
Pulsar mass&
1.29 \( \pm  \) 0.02 M\( _{\odot } \) &
1.17 \( \pm  \) 0.02 M\( _{\odot } \) \\
\hline 
\end{tabular}\par}

\caption{\label{params}The two degenerate best fits to the data. 
Solution 1 is more
likely, as the pulsar mass is more consistent with the current 
distribution of known
neutron star masses.}
\end{table}

\clearpage

\clearpage

\clearpage

\end{document}